\documentclass{article}
\usepackage[T1]{fontenc}
\usepackage[utf8]{inputenc}
\usepackage{ismir} 
\usepackage{amsmath,cite,url}
\usepackage{graphicx}
\usepackage{color}
\usepackage{amsthm}
\usepackage{mathbbol}
\usepackage{stmaryrd}
\usepackage[dvipsnames]{xcolor}
\usepackage{enumitem}
\usepackage{multirow}
\usepackage{stfloats}
\setlist[itemize]{noitemsep, nolistsep, leftmargin=*}

\title{Improved robustness in AI-generated music detection}

\multauthor
   {Emile Dugelay$^1$ \hspace{0.5cm} Thomas Barand$^1$ \hspace{0.5cm} Aurelien Laouar$^1$ \hspace{0.5cm} Baptiste Campeas$^1$}
   {{\bf  Darius Afchar$^2$ \hspace{0.5cm} Romain Hennequin$^2$ }\\
   $^1$Department of Computer Science, Ecole Centrale de Lyon, France\\
   $^2$Deezer Research, Paris, France\\
   {\tt\small \{dugelayemile, th.barand, aurelienlaouar, campeasb\}@gmail.com, research@deezer.com}
   }

\def\authorname{E. Dugelay, T. Barand, A. Laouar, B. Campeas, D. Afchar, R. Hennequin}

\begin{document}

\maketitle

\begin{abstract}

AI music generators leave predictable spectral artifacts determined by their architecture. Existing detectors exploit these artifacts with near-perfect accuracy on raw generated tracks, but their performance collapses under simple audio manipulations, such as speed modification or pitch shifting. We address this open robustness problem by introducing a frequency-scaling-invariant detection pipeline that aims to prevent this kind of attack by design. Our method maps audio onto a log-frequency axis via a log-STFT remapping.
A single learned cross-correlation filter, combined with max-pooling, provides shift invariance at inference time.
Training uses a hybrid loss that jointly supervises binary detection and artifact-peak localization, regularizing boundary weights.
Because robustness to speed change is built in by design, the detector is also interpretable: it outputs both a binary decision and an estimate of the applied speed-change factor.

\end{abstract}

\section{Introduction}\label{sec:introduction}

Audio GenAI tools like Suno, Udio, or Producer empower users with easy, prompt-based song creation. Although these tools open up creative possibilities, AI-generated tracks are widely used to flood streaming platforms as well as to drive fraudulent traffic as a malicious way to divert revenue from real artists \cite{DeezerNewsroom:01}. In the absence of transparent and reliable metadata, AI-generated music detection is essential to monitor content in the whole music recording industry (labels, distributors, music streaming platforms). Such detectors usually exploit spectral artifacts left by the generation process within the generated audio  \cite{DeezerAfchar:01}. These detectors prove effective on raw generated tracks, but perform poorly when the AI audio is manipulated with simple methods such as pitch shifting or resampling \cite{crosvila2025, afchar2025icassp}. Since these methods can be easily applied with free audio software or even directly in AI generation tools, this leaves vulnerabilities that fraudsters could easily exploit. Speed modification, in particular, is used as a creative tool in music genres such as nightcore \cite{Vinuela_Spedup} and is also widely used as a way to adapt music to video on platforms such as TikTok\cite{BBCSpedup:01}, so speed modification is very likely to be used on AI-generated content and to fool detectors. 
At the same time, keeping detectors lightweight enables fast inference and large-scale deployment and keeping them interpretable permits recourse in case of False Positives. In this paper, we introduce a frequency-scaling-invariant extension to the original Fourier strategy of \cite{DeezerAfchar:01} in order to make it robust to speed manipulation.

Our approach exploits a simple property: a frequency-scaling
transformation, such as the one induced by a speed modification, when viewed on a logarithmic frequency axis, becomes a
translation. This reframes robustness as a translation-invariance problem, for which we design a lightweight architecture whose invariance holds by construction rather than by training-time augmentation. The resulting models contain only a few thousand trainable parameters, match the state-of-the-art on clean audio, and remain near-perfect under speed modification attacks, a regime where baselines degrade substantially.

Our contribution is set on three axes. First, we formalize AI-music detection under speed modification attacks as a problem of log-frequency equivariance and translation invariance, and derive an architecture that satisfies both by construction. Second, we show that this architecture matches state-of-the-art detectors on unaltered audio while remaining robust under speed modification. Third, our detector is fully interpretable: beyond a binary decision, it outputs an estimate of the applied speed modification factor, enabling forensic analysis of the manipulation.
We publicly release the code so that anyone can replicate our experiments\footnote{\scriptsize{\url{github.com/thomas0barand/robust-deepfake-detector}}}.

\section{Related Work}\label{sec:related-work}


The detection of AI-generated music is an emerging MIR task sharing methodological overlap with composer identification and music auto-tagging~\cite{crosvila2025}, as well as deepfake detection \cite{wu2017asvspoof, mirsky2021creation, wang2020cnn}. Existing approaches can be categorized into two families.

\noindent\textbf{Content-based detection.}
The first family relies on learned representations of the musical content, either through end-to-end training or through classification on top of general-purpose embeddings. Rahman et al.~\cite{sonics2025} released SONICS, a large-scale dataset of Suno and Udio tracks generated from templated prompts, and proposed \textit{SpecTTTra}, a transformer handling long audio contexts by tokenizing spectrograms along both time and frequency.

Cros Vila et al.~\cite{crosvila2025} showed that simpler classifiers (SVMs, random forests, $K$-nearest neighbors) operating on CLAP audio embeddings \cite{wu2023large} can match or exceed SpecTTTra on their own Suno/Udio dataset, and benchmarked these against the commercial IRCAM Amplify detector. Their experiments revealed that all evaluated systems---including commercial ones---collapse under simple transformations: resampling the input to 22.05\,kHz alone is enough to flip the predicted label on a large fraction of AI samples. They further identified that such detectors rely on features across the 0 --12 kHz range, including both sub-500 Hz and above-10 kHz bands. It is consistent with the hypothesis that generation-specific cues concentrate in narrow frequency bands that post-processing can easily displace.

Similar detection approaches can be found in \cite{comanducci2025fakemusiccaps, zang2024singfake, afchar2025icassp, li2024audio, frohmann2025double, lopez2026ai, kim2025segment }. Arguably, these models are not interpretable due to their end-to-end nature.

\noindent\textbf{Explicit artifact detection.}
Other approaches design detectors around directly modeling generation artifacts, instead of embedding the whole musical content. 

Afchar et al.~\cite{DeezerAfchar:01} provided the first mathematical analysis of spectral artifacts in AI-generated music. They showed that transposed convolution layers---the standard upsampling primitive in neural audio decoders---produce predictable energy peaks at frequencies determined solely by network architecture (stride and hidden sampling frequency), independent of training data or learned weights. Controlled experiments confirmed this architectural-fingerprint hypothesis: the same architecture trained on different datasets (FMA, MTAT, MTG-Jamendo) or with different random seeds produced identical peak locations. This phenomenon is a frequency-domain instance of the well-known checkerboard artifact first analyzed in computer vision~\cite{odena2016deconvolution}, and noted in audio as ``upsampling artifacts'' by Donahue et al.~\cite{donahue2019wavegan} and Pons et al.~\cite{pons2020upsampling}; MelGAN~\cite{kumar2019melgan} explicitly chose kernel sizes to attenuate these effects.

Building on this characterization, Afchar et al.~\cite{DeezerAfchar:01}
proposed a lightweight detector that (i) time-averages the magnitude
spectrum, (ii) subtracts a piecewise-linear baseline to isolate peaks,
(iii) restricts analysis to a discriminative band
(\ 5--16\,kHz), and (iv) trains a logistic regression on the resulting feature, that we will refer to as \emph{fingerprint}, following the terminology of \cite{DeezerAfchar:01}.
The detector reached >99\% accuracy in their experiments, confirming that detecting the artifact was sufficient to detect AI generation.

\noindent\textbf{The Robustness Gap.}
Despite these recent advances, both families fail under simple frequency-scaling transformations applied
at inference. Cros Vila et al.~\cite{crosvila2025} documented this failure
across CLAP-based, transformer-based, and commercial systems, while Afchar
et al.~\cite{afchar2025icassp} reported a collapse of accuracy to 1--3\% under
pitch shift for their CNN-based detector.

The natural response is training-time data augmentation with speed-altered samples.
We instead propose to address the problem at the architectural level and extend the artifact-based detection line of \cite{DeezerAfchar:01} with a pipeline that is frequency-scaling-invariant by construction, preserving its interpretability and lightweight character.

\section{Problem Formulation }\label{sec:problem-formulation}

\subsection{Spectral Artifacts in AI-Generated Music}

Most AI music generators rely on transposed convolution layers to upsample latent embeddings into waveforms. As shown in~\cite{DeezerAfchar:01}, this upsampling process causes the signal spectrum to alias, producing spurious energy peaks at predictable frequencies. For a single deconvolution layer of stride $k$ and hidden sampling frequency $f_s$, these artifacts form a finite Dirac comb of $\lfloor k/2 \rfloor + 1$ peaks evenly spaced by $f_s$. When $L$ such layers are stacked, the per-layer combs combine into a set $\mathcal{P}$ of peaks with a \textit{fractal-like} structure: $\mathcal{P}$ contains every integer combination of the layers' hidden sampling frequencies, weighted by their respective strides. This fingerprint is determined by the network architecture alone, independent of training data or learned weights. We refer to the representation of such fingerprints learned by our model as a \textit{fakeprint}.

\subsection{Detection Pipeline}

Given an audio signal $s(t)$, we model the detection system as a two-stage pipeline. An artifact extractor $A$ first maps $s$ to a representation $A(s)$ that highlights the spectral peaks described by $\mathcal{P}$. A binary classifier $D$ then operates on this representation to decide whether the signal is AI-generated:
\begin{equation}
    s(t)
    \;\xrightarrow{\;A\;}
    A(s)
    \;\xrightarrow{\;D\;}
    D(A(s)) \in \{0, 1\}.
\end{equation}
This factored design is intentional: by separating artifact extraction from classification, each component can be studied and optimized independently.

\subsection{Signal Transformations and Robustness}

A basic attack against such a detector is to apply a post-processing transformation to the generated audio before it is submitted for analysis. We focus on the very common \textit{speed modification} \cite{Vinuela_Spedup,BBCSpedup:01} (note that we address direct speed modifications with their induced frequency alteration, not time-stretching).

\noindent\textbf{Speed modification} by a factor $\alpha > 0$ consists of resampling the signal as $T_{\alpha}:s(t) \mapsto s(\alpha t)$, which scales the temporal signal by $\alpha$ and the frequencies by $\frac{1}{\alpha}$.\\

Here, we study theoretically how the detection pipeline behaves under ideal spectral scaling $T_\alpha$.

Under $T_\alpha$, the full pipeline becomes:
\begin{equation}
    s(t)
    \;\xrightarrow{\;T_\alpha\;}
    s(\alpha t)
    \;\xrightarrow{\;A\;}
     A \circ T_\alpha(s)
    \;\xrightarrow{\;D\;}
    D \circ A \circ T_\alpha (s).
\end{equation}

\begin{figure*}[t]
    \centering
    \includegraphics[width=\textwidth]{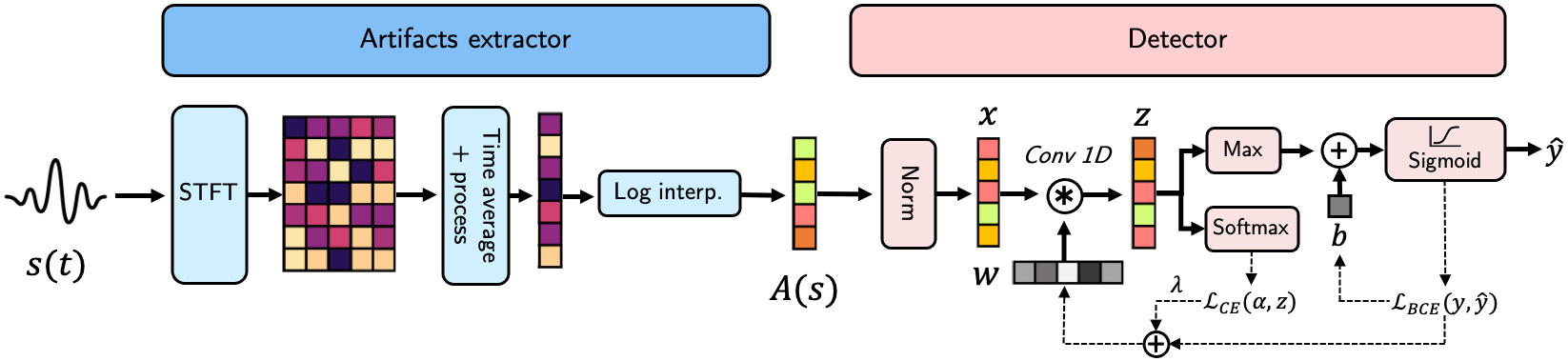}
    \caption{Overview of the proposed detection pipeline. The artifact extractor maps audio $s(t)$ to a log-frequency fakeprint $A(s)$ via STFT, time-averaging, and logarithmic interpolation. The detector normalizes (batch) $A(s)$, into $\mathbf{x}$ and computes its cross-correlation $\mathbf{z}$ with learned weights $\mathbf{w}$ via 1D convolution. The output $\hat{y}$ is obtained by applying a sigmoid to $\max(\mathbf{z}) + b$, where $b$ is a trainable bias. Training jointly minimizes a BCE loss $\mathcal{L}_{\mathrm{BCE}}(y, \hat{y})$ on the binary AI/human label and, weighted by $\lambda$, a cross-entropy loss $\mathcal{L}_{\mathrm{CE}}(\alpha, \mathbf{z})$ that supervises the peak location in $\mathbf{z}$.}
    \label{fig:pipeline}
\end{figure*}

\subsection{Equivariance and Invariance Objectives}

We formalize the detection task through the notions of equivariance and invariance.

\noindent\textbf{Log-frequency equivariance of \textit{A}.}
A speed change transformation $T_\alpha : s(t) \mapsto s(\alpha t)$ acts on the spectrum by scaling all frequencies by a factor $\frac{1}{\alpha}$. When the spectrum is represented on a logarithmic frequency axis $\xi = \log f$, this multiplicative factor becomes an additive translation by $-\log \alpha$, as $\log \frac{f}{\alpha} = \log(f) - \log{\alpha}$. We therefore seek an extractor $A$ satisfying:
%
\begin{equation}
    A \circ T_{\alpha}(s)(\xi) = A(s)({\xi - \log\alpha}), \quad \forall s,\, \forall \alpha > 0,
\end{equation}

where the right-hand side denotes the feature map of $s$ translated by $\log\alpha$ along the log-frequency axis. In other words, scaling the input signal by $\alpha$ produces a rigidly shifted version of the extracted features. The artifact pattern moves, but its structure is preserved.

\noindent\textbf{Invariance of \textit{D}.}
The classifier $D$ should be insensitive to such translations, i.e.:
\begin{equation}
    D \circ A(s)({\xi + k}) = D \circ A(s)(\xi), \quad \forall s,\, \forall k .
\end{equation}
These two objectives are complementary: if $A$ satisfies log-frequency equivariance, the robustness of the full pipeline $D \circ A$ reduces to requiring $D$ to be invariant to translations along the log-frequency axis.

\section{Proposed Method}\label{sec:method}

We now describe the concrete implementation of the two-stage pipeline introduced in Section~\ref{sec:problem-formulation}. The artifact extractor $A$ is engineered to satisfy log-frequency equivariance, the detector $D$ is built to be invariant to such translations, so that the full pipeline $D \circ A$ remains robust to time-frequency scaling $T_\alpha$. An overview of both components is illustrated in Figure~\ref{fig:pipeline}.

\subsection{Artifact Extractor}

Given a signal $s(t)$, the artifact extractor $A$ produces a 1D log-frequency representation $A(s) \in \mathbb{R}^F$, with $F$ the number of log-frequency bins, that highlights the spectral peaks described by $\mathcal{P}$. It proceeds in three steps: time-frequency analysis, time averaging, and lower-hull subtraction.

\noindent\textbf{Time-frequency analysis.}
In line with \cite{DeezerAfchar:01}, the \textit{Short-Time Fourier Transform} (STFT) is applied to the audio signal $s(t)$, yielding a spectrogram with linearly spaced frequency bins.

\noindent\textbf{Time averaging and log remapping.}
The spectrogram is collapsed to a 1D spectrum by averaging over the time dimension. The STFT spectrum $\bar{S}$ is then linearly interpolated onto $F$ geometrically spaced bins:
\begin{equation}
    \bar{S}_{\log}[\xi_k] = \operatorname{interp}(\bar{S})\!\left(
    f_{\min} \cdot 2^{k/B'}\right), \quad k \in \llbracket 0,\, F-1 \rrbracket,
\end{equation}
where $B'$ is the target log-frequency resolution.

\noindent\textbf{Lower-hull subtraction.}
Again, following \cite{DeezerAfchar:01}, we subtract a piecewise-linear baseline $H(\bar{S}_{\log})$ to isolate the artifact peaks in fakeprints:
\begin{equation}
    A(s)[\xi] = \max\!\left(\bar{S}_{\log}[\xi] - H(\bar{S}_{\log})[\xi],\;0\right).
\end{equation}

$A(s)$ retains only peaks above the smooth baseline, isolating the artifacts from $\mathcal{P}$. Under an ideal frequency scaling $T_\alpha$, the whole fakeprint shifts rigidly by $\log\alpha$ along $\xi$, as illustrated in Figure~\ref{fig:cross-correlation-fakeprint-w}.

\subsection{Detector}\label{sec:detector}
 
The detector $D$ takes as input the normalized fakeprint $\mathbf{x} \in \mathbb{R}^F$ and produces a binary prediction $\hat{y} \in [0,1]$ together with a lag estimate $\hat{\alpha}$. It comprises three components: batch normalization, a 1D convolution with a single learned filter $\mathbf{w} \in \mathbb{R}^F$, and a classification head.

\noindent\textbf{Normalization.}
The fakeprint $A(s)$ is first clamped (prevents outlier bins from dominating) and batch-normalized (standardizing feature distributions across the batch).

\begin{figure}[t]
    \centering

    \includegraphics[width=\columnwidth]{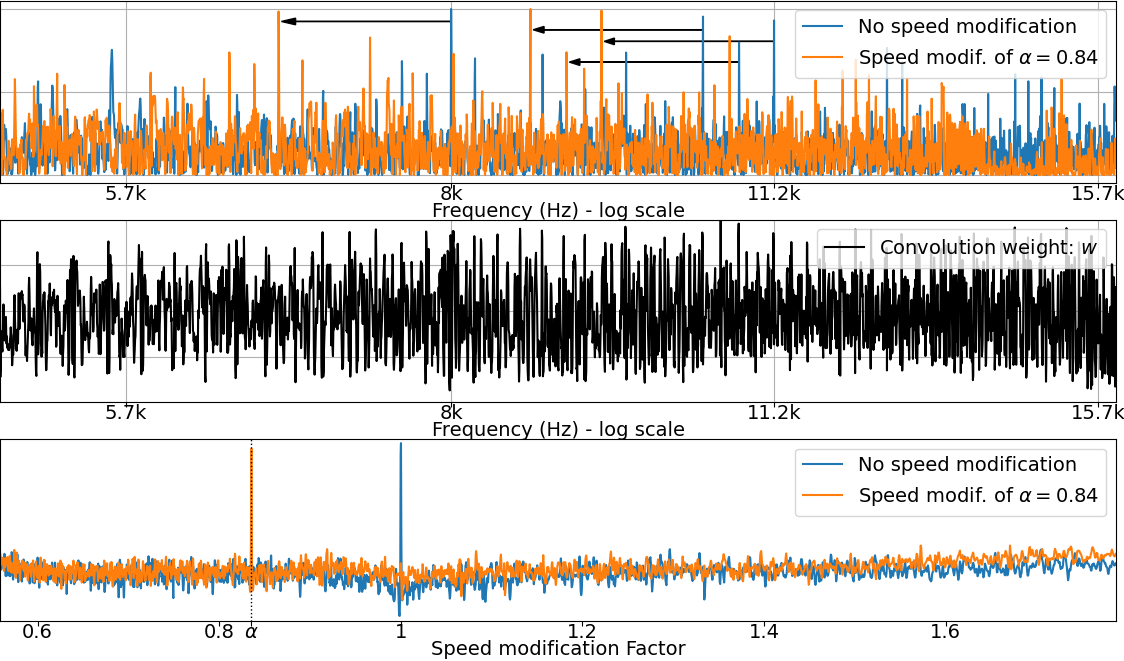}
    \caption{Illustration of the learned detector on a generated track and the same track after a speed modification of $\alpha = 0.84$.
    \textbf{Top:} fakeprints of both tracks. A speed change shifts the artifact pattern rigidly along the log-frequency axis (arrows).
    \textbf{Middle:} learned convolution weights $\mathbf{w}$. Peaks emerge at the characteristic locations of the unshifted artifact pattern --- visible for instance around $8\,\mathrm{kHz}$ and $11.2\,\mathrm{kHz}$, aligned with the blue fakeprint above.
    \textbf{Bottom:} cross-correlation between $\mathbf{w}$ and the two fakeprints. Each curve exhibits a sharp peak whose position directly indicates the applied speed-change factor: $1.0$ for the unmodified track and $\alpha$ for the manipulated one.}

    \label{fig:cross-correlation-fakeprint-w}
\end{figure}

\noindent\textbf{Cross-correlation via 1D convolution.}
The core of the detector is a 1D convolution between $\mathbf{x}$ and $\mathbf{w}$ leading to $\mathbf{z} \in \mathbb{R}^F$.
Each entry $z_k$ is the inner product of $\mathbf{x}$ shifted by $k - \left\lfloor \frac{F}{2} \right\rfloor$ bins with $\mathbf{w}$, which is equivalent to the cross-correlation between $\mathbf{x}$ and the time-reversed filter $\mathbf{w}$, evaluated at lag $k - \left\lfloor \frac{F}{2} \right\rfloor$. A peak at $k^* = \operatorname{argmax}_k z_k$ indicates that the fakeprint best aligns with $\mathbf{w}$ when shifted by $k^* - \left\lfloor \frac{F}{2} \right\rfloor$ bins, as shown in Figure~\ref{fig:cross-correlation-fakeprint-w}.

\noindent\textbf{Classification head.}
The detection score is obtained by applying a sigmoid to the cross-correlation maximum plus a trainable bias $b$:
\begin{equation}
    \hat{y} = \sigma\!\left(\max_k z_k + b\right).
\end{equation}
The bias $b$ controls the global detection threshold independently of the shape of $\mathbf{w}$.

\noindent\textbf{Shift invariance.}
The max-pooling over $\mathbf{z}$ makes the classification head invariant to translations of the fakeprint. If $\mathbf{x}$ is shifted by $\Delta$ bins, every entry of $\mathbf{z}$ shifts cyclically by $\Delta$, so $\max_k z_k$ remains unchanged.


\subsection{Training objective}
\label{sec:training_obj}

The model is trained with a hybrid loss combining a Binary Cross-Entropy (BCE) classification term and a Cross-Entropy (CE) lag-regression term:
\begin{equation}
    \mathcal{L} = \mathcal{L}_{\mathrm{BCE}}(y,\hat{y})
                + \lambda\,\mathcal{L}_{\mathrm{CE}}(\alpha, \mathbf{z}),
\end{equation}
where the CE term is computed only on AI-generated samples (for which a ground-truth lag $\alpha$ is available). The CE target is a one-hot vector with a $1$ at position $\lfloor F/2 + \log_2(\alpha)\cdot B'\rceil$, i.e.\ the bin index that the peak of $\mathbf{z}$ should occupy.
 
\noindent\textbf{Gradient analysis.}
The two terms act on $\mathbf{w}$ through distinct gradient signals. Let $k^* = \operatorname{argmax}_k z_k$ denote the current peak position. The BCE gradient updates $\mathbf{w}$ through a single cross-correlation slice:
\begin{equation}
    \frac{\partial \mathcal{L}_{\mathrm{BCE}}}{\partial w_f}
    = (\hat{y} - y)\cdot x_{k^*+f-\left\lfloor \frac{F}{2} \right\rfloor},
\end{equation}

while the CE gradient aggregates over all $F$ lags:
\begin{equation}
    \frac{\partial \mathcal{L}_{\mathrm{CE}}}{\partial w_f}
    = \sum_{k=0}^{F-1} \bigl(p_k - \mathbf{1}[k = k_\alpha]\bigr)\cdot x_{k+f-\left\lfloor \frac{F}{2} \right\rfloor},
\end{equation}

where $p_k = \operatorname{softmax}(\mathbf{z})_k$ and $k_\alpha$ is the target lag index. Unlike the BCE gradient, which is non-zero only at the single winning lag $k^*$, the CE gradient distributes across all weights $w_f$ through all the lags, ensuring a dense and uniform gradient signal across all frequency bins. We show in Section \ref{sec:results} that the addition of the CE term has a slightly positive impact on the performance of detection of the proposed system.

Because the BCE gradient touches $w$ only through the slice at $k^*$, weights $w_f$ whose receptive field index $k^*+f - \lfloor F/2 \rfloor$ lies in the zero-padded region receive a near-zero gradient. The CE term, by summing over all $k$, provides a dense gradient signal to every weight including boundary ones, keeping the learned weights balanced across all frequency bins rather than collapsing to zero at the boundaries (which we observed experimentally).

\subsection{Interpretability}
\label{sec:interpretability}
By construction, the learned $\mathbf{w}$ acts as a \textit{reference fakeprint}: it encodes the expected log-frequency pattern of artifacts for the target architecture. At inference time, the cross-correlation $\mathbf{z}$ measures the similarity between the fakeprint $\mathbf{x}$ and $\mathbf{w}$ at every possible log-frequency lag.

This yields a dual output. The maximum value of $\mathbf{z}$, passed through the sigmoid, drives the binary detection score. The position of that maximum, $k^*$  locates the lag at which $\mathbf{x}$ and $\mathbf{w}$ best align, and directly recovers the speed-change factor applied to the input.
Figure~\ref{fig:cross-correlation-fakeprint-w} illustrates this behaviour on a representative example.

\section{Experimental Setup}\label{sec:experimental-setup}

\subsection{Dataset}

To train our network to distinguish between AI-generated and genuine audio, we compiled a dataset of 10,000 audio tracks. Each track was truncated to retain only the first 30 seconds of audio. The dataset is evenly balanced, 5,000 AI-generated tracks sourced from Suno v5 and 5,000 genuine tracks sampled from the FMA-small dataset~\cite{DBLP:journals/corr/BenziDVB16}. In addition to Suno v5, we trained and evaluated our method on Suno v3.5 and Udio v120, with 5,000 tracks for each music generator both taken from the Sonics dataset~\cite{sonics2025}.

To evaluate the model's robustness against temporal alterations, we established two distinct training subsets:

\begin{itemize}
    \item \textbf{"Clean" Dataset:} Consists of the original audio tracks without any modifications.
    \item \textbf{"Attack" Dataset:} Consists of tracks manipulated with speed modification. For training, speed-change factors were uniformly sampled from a discrete set of bin shifts in [-990, +990] bins, which corresponds to an approximate speed multiplier range of $\times$0.7 to $\times$1.4. The sampling is restricted to discrete bin-aligned values so that each applied speed factor corresponds exactly to an integer shift in the log-frequency representation. This ensures an unambiguous ground-truth target for $\mathcal{L}_{\mathrm{CE}}$. At test time, the speed-change factors were sampled uniformly and continuously from the range $\times$0.7 to $\times$1.4, without any constraint of alignment to discrete bin shifts, so that the evaluation reflects realistic speed manipulations rather than bin-aligned ones. The specific speed change factor applied to each track was recorded and saved alongside its extracted features.
\end{itemize}

\subsection{Feature Extraction and Preprocessing}

Our data comes from two sources with different native sampling rates: the \textsc{sonics} dataset\cite{sonics2025} at $16$\,kHz, and Suno v5 recordings at $44{,}100$\,Hz. We keep each source at its native rate to avoid resampling artifacts. 

The network takes as input the feature we refer to as the \emph{fakeprint}, computed in a manner similar to \cite{DeezerAfchar:01}: we use a Fourier transform frame of size $N_{\text{fft}} = 2^{14}$. It is cropped to a frequency band that depends on the sampling rate: $1$--$7$\,kHz for the \textsc{sonics} data and $5$--$16$\,kHz for Suno v5. The frequency axis is then mapped to a log-scale by linear interpolation onto $1920$ bins per octave.

\subsection{Training Procedure}

The network was trained using a $10\%$ validation split for a maximum of $50$ epochs with a batch size of $64$. We utilized the Adam optimizer, initialized with a learning rate of $0.001$ and a weight decay of $10^{-5}$. The cross-entropy loss is modulated by a tuning hyperparameter, $\lambda$. Thanks to the offline caching of the fakeprints and the compact architecture, training completes in under 5 minutes on a consumer-grade GPU.

\section{Results \& Discussion }\label{sec:results}

We evaluate our method across two main axes: \textbf{(1)} Detection performance on unattacked audio compared to two baselines: our implementation  of Afchar et al.~\cite{DeezerAfchar:01} and the SpecTTTra-$\alpha$ (best performing model of \cite{sonics2025}), \textbf{(2)} Robustness to speed change attacks across all model variants.

Finally, we show that the robustness to speed modification of our system also induces a limited form of robustness to pitch-shift modifications.

\subsection{Effect of the CE speed factor classification loss}\label{sec:gradient}

Table~\ref{tab:lambda} shows the effect of the auxiliary lag-regression loss weight $\lambda$ on our model, trained and evaluated on speed-modified Suno v5 recordings (corresponding to the \textit{Attack}/\textit{Attack} condition of Table~\ref{tab:merged-results}). Setting $\lambda{=}0$ (BCE only) already yields strong results thanks to the log-frequency design, but boundary weights collapse toward zero as discussed in Section~\ref{sec:training_obj}.

\begin{table}[h]
\centering
\setlength{\tabcolsep}{4pt}
\renewcommand{\arraystretch}{1.1}
\small
\begin{tabular}{|c || c c c|}
\hline
$\lambda$ & AUC & F1 & Prec. \\
\hline
0.00 & 0.996 & 0.963 & \textbf{0.991} \\
0.05 & \textbf{0.997} & \textbf{0.986} & 0.989 \\
0.10 & \textbf{0.997} & 0.979 & \textbf{0.991} \\
\hline
\end{tabular}
\caption{Tuning on $\lambda$ for \textbf{Our Method} evaluated on attacked audio. Best value highlighted in \textbf{bold}.}
\label{tab:lambda}
\end{table}

By looking especially at F1, adding the CE loss with $\lambda>0$ provides a dense gradient signal that regularizes $\mathbf{w}$ and consistently improves all metrics showing the importance of this extra loss. We can see the performance is not very sensitive to the value of $\lambda$: we observe a slight decrease of F1 and AUC for $\lambda>0.05$ which could be explained by the domination of the CE loss over the main BCE loss which aims at solving the classification task we evaluate on.

In the remainder of the paper, we then report only performance of our system for $\lambda{=}0.05$.

\subsection{AI Detection and Scale Invariance}

\noindent\textbf{Performance on unattacked audio:} Table~\ref{tab:merged-results} compares our model against SpecTTTra-$\alpha$~\cite{sonics2025} and Afchar et al.~\cite{DeezerAfchar:01} under all combinations of training and test conditions. When trained and tested on clean, unmodified audio (first block), our model nearly matches the baseline despite containing few trainable parameters, confirming that the log-frequency cross-correlation is a sufficient statistic for artifact detection on unperturbed signals. SpecTTTra-$\alpha$ yields poor results on Suno v5 primarily because it was not trained on the Suno v5 data.

\begin{table*}[t]
\centering
\renewcommand{\arraystretch}{0.95}
\footnotesize
\begin{tabular*}{\textwidth}{@{\extracolsep{\fill}}|l l l|| c c c || c c c || c c c|}
\hline
 & & & \multicolumn{3}{c||}{Suno v3.5} & \multicolumn{3}{c||}{Suno v5} & \multicolumn{3}{c|}{Udio v120} \\
\cline{4-12}
\textbf{Train} & \textbf{Test} & \textbf{Model} & AUC & F1 & Prec. & AUC & F1 & Prec. & AUC & F1 & Prec. \\
\hline
\multirow{3}{*}{Clean} & \multirow{3}{*}{Clean}
 & Afchar et al. & \textbf{1.000} & \textbf{.998} & \textbf{.996} & \textbf{1.000} & \textbf{.998} & \textbf{.999} & \textbf{1.000} & \textbf{.997} & \textbf{.994} \\
 & & SpecTTTra-$\alpha$\textsuperscript{\dag} & .997 & .965 & .932 & .714 & .360 & .793 & .999 & .965 & .932 \\
 & & Our Method & .999 & .995 & .990 & \textbf{1.000} & \textbf{.998} & \textbf{.999} & .996 & .990 & .982 \\
\hline
\multirow{2}{*}{Attack} & \multirow{2}{*}{Clean}
 & Afchar et al. & .968 & .919 & .883 & .946 & .885 & .849 & .939 & .895 & .839 \\
 & & Our Method & \textbf{1.000} & .995 & .992 & .999 & .993 & .995 & .988 & .965 & .968 \\
\hline
\hline
\multirow{3}{*}{Clean} & \multirow{3}{*}{Attack}
 & Afchar et al. & .893 & .004 & .250 & .745 & .008 & .400 & .864 & .021 & .556 \\
 & & SpecTTTra-$\alpha$\textsuperscript{\dag} & .780 & .552 & .815 & .692 & .302 & .687 & .810 & .589 & .848 \\
 & & Our Method & .977 & .754 & .979 & .986 & .732 & \textbf{.993} & .932 & .574 & \textbf{.950} \\
\hline
\multirow{2}{*}{Attack} & \multirow{2}{*}{Attack}
 & Afchar et al. & .910 & .782 & .848 & .817 & .675 & .768 & .918 & .842 & .825 \\
 & & Our Method & \textbf{.999} & \textbf{.983} & \textbf{.989} & \textbf{.997} & \textbf{.986} & .989 & \textbf{.964} & \textbf{.855} & \textbf{.950} \\
\hline
\end{tabular*}
\caption{Detection performance for all combinations of training and test conditions (\textit{Clean}: unmodified audio; \textit{Attack}: random speed modification, bin-aligned factors for training, continuous non-bin-aligned factors for test, see Section~\ref{sec:experimental-setup}).\\\textsuperscript{\dag}: Evaluated as released, not retrained. Best value per metric and test condition in \textbf{bold}.}
\label{tab:merged-results}
\end{table*}

\noindent\textbf{Robustness under speed modification attacks:}
The bottom blocks of Table~\ref{tab:merged-results} report performance for all model variants when models are evaluated on attacked audio, with speed factors drawn continuously (not bin-aligned).

 The Afchar et al. baseline, which operates on a linearly spaced frequency axis, collapses under these transformations (AUC $\approx 0.81$ and F1 $\approx 0.67$ on Suno v5). SpecTTTra-$\alpha$, evaluated as released, degrades similarly. Our pipeline maintains high performance, validating the theoretical guarantees derived in Section~\ref{sec:problem-formulation}. It achieves an AUC of $0.997$ and a F1-Score of $0.986$ under attack, outperforming both evaluated baselines.
 
Beyond detection, our architecture also enables recovering the applied speed-change factor from the cross-correlation peak location. On Suno v5, this estimate reaches an MAE of $4 \times 10^{-4}$.

 \begin{figure}[h]
    \centering
    \includegraphics[width=\columnwidth]{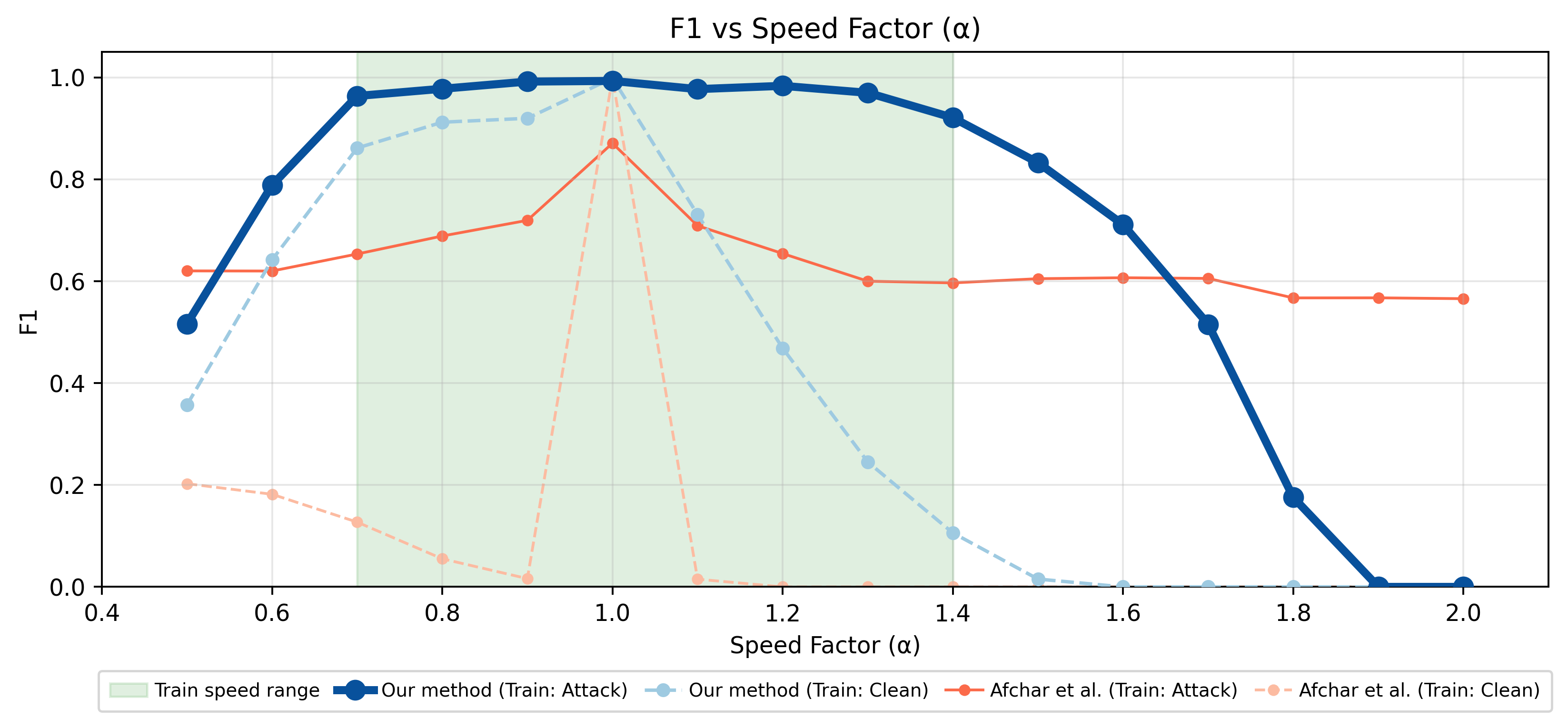}
    \caption{\textbf{F1-Score vs Speed Factor ($\alpha$).} Evaluation of model generalization across a range of speed factors on Suno v5, for Our method and Afchar et al.. While our model was trained on the interval $[0.7, 1.4]$ (shaded), it maintains an F1-score $> 0.8$ up to a speed factor of $1.5$.}
    \label{fig:f1_vs_af}
\end{figure}

\noindent\textbf{Generalization to unseen speed modification factors:}
To further highlight the inherent robustness of our architecture, we emphasize the model's performance when trained exclusively on unaltered tracks and evaluated against speed modification attacks (\textit{Clean}/\textit{Attack} block of Table~\ref{tab:merged-results}). Even without exposure to temporally altered data during training, our model remains highly performant (AUC stays above $0.93$ on all datasets), demonstrating generalization capacity by design. In addition, Figure~\ref{fig:f1_vs_af} demonstrates that our approach remains robust to track alterations, exhibiting strong generalization even when subjected to speed factors just outside the training distribution, while Afchar et al. baseline remains below at every speed factor $\neq 1$.

\noindent\textbf{Robustness to Pitch Shifting:} The most general form of pitch-shifting implementation is based on time-stretching (usually performed by a phase vocoder) followed by a resampling of the waveform to bring it back to its original duration. Here, we use the Rubber Band Library\cite{rubberband2024} to generate high-quality pitch-shifted samples. The time-stretching phase will affect the artifacts' strength, but will not change their position, and detectors are expected to retain partial robustness to time-stretching, as shown in \cite{DeezerAfchar:01}. The resampling will move all artifacts and will result in a drop of the performance of detectors, similarly as we observe in Table~\ref{tab:merged-results}. The speed change invariance of our proposed method is naturally designed to compensate the effect on frequency of this second phase.
The results of the detectors applied on pitch-shifted signals are shown in Table~\ref{tab:pitch_shift}. We do observe a limited robustness to pitch-shifting in our detector, particularly in maintaining high precision/AUC and overall separability. The main effect of pitch-shifting is to decrease the detection score of our method, but the very high AUC shows a high discriminative power, and other metrics could possibly be optimized by recalibrating the decision threshold, or optionally retraining on additional pitch-shifted samples.

\begin{table}[h]
\centering
\renewcommand{\arraystretch}{1.1}
\small
\resizebox{\columnwidth}{!}{%
\begin{tabular}{|l|| c c c | c c c | c c c|}
\hline
 & \multicolumn{3}{c|}{Suno v3.5} & \multicolumn{3}{c|}{Suno v5} & \multicolumn{3}{c|}{Udio v120} \\
\cline{2-10}
\textbf{Model} & AUC & F1 & Prec. & AUC & F1 & Prec. & AUC & F1 & Prec. \\
\hline
Afchar et al. & 0.794 & \textbf{0.732} & 0.624 & 0.771 & \textbf{0.720} & 0.697 & 0.768 & \textbf{0.720} & 0.572 \\
SpecTTTra-$\alpha$ & 0.674 & 0.399 & 0.774 & 0.700 & 0.347 & 0.686 & 0.764 & 0.557 & \textbf{0.821} \\
Our Method & \textbf{0.920} & 0.439 & \textbf{0.931} & \textbf{0.963} & 0.279 & \textbf{0.987} & \textbf{0.811} & 0.707 & 0.767 \\
\hline
\end{tabular}%
}
\caption{Detection performance on audio altered with pitch shift. We evaluate the same augmentation-trained models as in Table~\ref{tab:merged-results} (\textit{Attack}-trained rows).}
\label{tab:pitch_shift}
\end{table}

\subsection{Limits}\label{sec:limits}
A common alternative to architectural invariance is training-time augmentation with speed-altered or pitch-shifted samples. We did not pursue this route, as it trades theoretical guarantees for empirical coverage of a discrete set of attack parameters and must be revisited for each new transformation. Our pipeline is invariant to frequency scaling by construction, as confirmed by the zero-shot robustness in Table~\ref{tab:merged-results}: trained on unaltered audio alone, the model still reaches AUC $0.986$ under attack.

Two limitations remain: first, we targeted a single attack type. Other common manipulations (equalization, dynamic range compression, codec compression) remain untested, and pitch-shifting (Table~\ref{tab:pitch_shift}), which combines resampling and time-stretching, exposes the limits of our representation: time-stretching is not equivariant under it, nor would be a non-uniform warp scaling frequency bands by different factors.

Second, training was restricted to three generative models, so the model does not generalize to other generators. This limitation is not specific to our approach: SpecTTTra-$\alpha$, the detector of Afchar et al., and other state-of-the-art systems exhibit the same behavior, as each generative architecture leaves distinct spectral artifacts that must be learned from data. Retraining is therefore required for broad coverage across all systems, but the lightweight design of our model makes this process particularly inexpensive.

\section{Conclusion}\label{sec:conclusion}

This paper addresses a critical vulnerability in current AI music detectors: their drastic accuracy drop under simple frequency-scaling manipulations such as speed modification. We introduced a mathematically grounded, lightweight pipeline that is frequency-scaling-invariant by design: mapping audio to a log-frequency axis yields shift equivariance, while a 1D cross-correlation filter with max-pooling guarantees shift invariance at inference. 

Our model matches state-of-the-art performance on clean audio and reaches an F1-Score up to 0.986 under speed-change attacks, outperforming the baseline while recovering the speed-change factor for forensics, with partial robustness to pitch-shifting.

\clearpage

\section{Ethics Statement}\label{sec:ethics}

The detector proposed in this paper is designed to support content
monitoring on streaming platforms, where AI-generated tracks are
increasingly used to extract fraudulent royalties at scale. Reliable
detection contributes to a fairer distribution of revenue toward human
artists and to greater transparency for listeners.

Despite these benefits, our system and others carry risks that necessitate careful deployment. False positives pose an asymmetric threat, as misidentifying human-made tracks can cause irreparable financial and reputational harm; thus, detection scores should be treated as auxiliary signals within a human-in-the-loop framework rather than a sole basis for automated enforcement. Furthermore, because the boundary between human and AI-generated music is increasingly porous due to widespread AI-assisted production \cite{crosvila2025}, binary classification risks unfairly penalizing legitimate hybrid works. We therefore emphasize the need for clear communication regarding which specific forms of AI involvement the model flags to ensure transparency for both creators and listeners.

\bibliography{refs}

\end{document}